\documentclass[10pt, conference]{IEEEtran}
\usepackage{cite}

\usepackage{balance}
\usepackage[cmex10]{amsmath}
\usepackage{amsthm}
\usepackage{amssymb}
\usepackage{tabulary}
\usepackage{booktabs}
\usepackage{bm}
\usepackage{pgfplots}
\usepackage{caption}
\usepackage{subcaption}
\usetikzlibrary{patterns}
\usepackage{enumerate}
\usepackage[subnum]{cases}
\usepackage[overload]{empheq}

\usepackage{makecell}

\usepackage[]{algorithm}
\usepackage{algpseudocode}

\usepackage{etoolbox}

\usepackage{graphicx}
\usepackage{amsmath,amsthm,bm,mathrsfs}

\renewcommand{\IEEEproofindentspace}{0\parindent}

\renewcommand\IEEEkeywordsname{Keywords}

\title{Privacy-Preserving Mechanisms for Parametric Survival Analysis with Weibull Distribution}

\author{
\IEEEauthorblockN{Th\^ong T. Nguy\^en}
\IEEEauthorblockA{Nanyang Technological University\\s140046@ntu.edu.sg}\\
\and
\IEEEauthorblockN{Siu Cheung Hui}
\IEEEauthorblockA{Nanyang Technological University\\
asschui@ntu.edu.sg}\\
}

\def\header{\vspace{1.5mm} \noindent}

\theoremstyle{plain}

\newtheorem{theorem}{Theorem}
\newtheorem{lemma}{Lemma}

\newtheorem{definition}{Definition}

\pgfplotsset{compat=newest}

\pgfplotsset{soldot/.style={color=black,only marks,mark=*}} \pgfplotsset{holdot/.style={color=black,fill=white,only marks,mark=*}}

\colorlet{shadecolor}{blue!20}
\begin{document}

\maketitle

\begin{abstract}
Survival analysis studies the statistical properties of the time until an event of interest occurs. It has been commonly used to study the effectiveness of medical treatments or the lifespan of a population. However, survival analysis can potentially leak confidential information of individuals in the dataset. The state-of-the-art techniques apply ad-hoc privacy-preserving mechanisms on publishing results to protect the privacy. These techniques usually publish sanitized and randomized answers which \emph{promise} to protect the privacy of individuals in the dataset but without providing any formal mechanism on privacy protection. In this paper, we propose private mechanisms for parametric survival analysis with Weibull distribution. We prove that our proposed mechanisms achieve differential privacy, a robust and rigorous definition of privacy-preservation. Our mechanisms exploit the property of local sensitivity to carefully design a utility function which enables us to publish parameters of Weibull distribution with high precision. Our experimental studies show that our mechanisms can publish useful answers and outperform other differentially private techniques on real datasets.
\end{abstract}

\vspace{2mm}

\section{Introduction}

Survival analysis \cite{miller2011survival, klein2005survival} focuses on modeling the time to an event of interest. It is used in clinical trials to study the efficacy of a treatment \cite{kantoff2010overall,fleming2000survival, marubini2004analysing}. It is also used to study the lifespan of a population, the time until a machine fails, etc.  However, there are concerns about the privacy risk in survival analysis  as it may potentially leak confidential information to the public. For example, it is possible that the results from a study about the survival of HIV patients may leak the identity of individuals participated in the study. 
Moreover, many countries publish survival analysis results of the population to the public. But, any breach of privacy from these results can be a disaster on a massive scale.

The survival function and hazard function are the two most important focuses in survival analysis. The survival function $S(t)$ is the probability of survival at least to time $t$, while the hazard function $h(t)$ is the risk of death at time $t$. There are three kinds of models used in survival analysis: (1) non-parametric models are used to estimate $S(t)$ and $h(t)$ directly from data without any assumption on data distribution, e.g., the Kaplan-Meier estimator; (2) parametric models are used to estimate $S(t)$ and $h(t)$ under the assumption that the data follow a parametrized distribution such as Weibull distribution; and (3) semi-parametric models are used to estimate the effect of variables on survival, e.g., Cox regression. Different models will need different privacy-preserving techniques. In this paper, we focus on  privacy protection for parametric models only. In particular, we focus only on the parametric model with Weibull distribution, which is the most common model used in parametric survival analysis. In this model, $S(t)$ and $h(t)$ are parametrized by two parameters of Weibull distribution, namely the shape parameter $p$ and the scale parameter $\lambda$. The state-of-the-art technique to protect privacy in parametric survival analysis is the work from O'Keefe \emph{et al.} \cite{o2012confidentialising} which proposed to randomly sample 95\% of the datasets for estimation. The estimated parameters of the Weibull distribution are then rounded to protect the privacy of individuals in the datasets. However, the proposed technique has no rigorous control on how much confidential information may be leaking to the public. As far as we know, this is the only work focusing on privacy-preserving techniques for parametric survival analysis.

Our work aims to propose mechanisms for estimating parametric models under the formal protection of differential privacy, which is the state-of-the-art privacy-preserving technique. Differential privacy is a mathematical definition of privacy which guarantees that the addition or removal of any individual in the dataset cannot gradually change the probability distribution of the output. Therefore, it prevents anyone including the adversary from learning about the individuals in the dataset. Let $D$ be a dataset, and each row in the dataset $D$ is a record of an individual. The distance between two datasets $D$ and $D^\prime$,  denoted as $d(D, D^\prime)$, is the number of different rows between $D$ and $D^\prime$. And two datasets are \emph{adjacent} to each other if the distance between them is equal to $1$. The mathematical definition of differential privacy is as follows.
\begin{definition}[Differential privacy \cite{dwork2006calibrating}]
	A mechanism (or function) $\mathcal M$ is $\epsilon-$differentially private if for any pair of adjacent datasets ${D}$ and ${D}^\prime$, and for any value $x$:
	\begin{align}
	\text{Pr} \left[ \mathcal M ({D}) = x \right] \leq 
	e^\epsilon \cdot \text{Pr}\left[  \mathcal M({D}^\prime) = x \right],  \label{dp}
	\end{align} 
where $\epsilon$ is the privacy budget of  $\mathcal M$. 
\end{definition}
\noindent The privacy budget $\epsilon$ is a quantitative measurement on how much \emph{privacy} of the individuals in the dataset is consumed by $\mathcal M$ when $\mathcal M(D)$ is published and available to everyone.

In this paper, we propose a two-step approach to estimate the parameters $p$ and $\lambda$ of Weibull distribution privately. First,  we propose \emph{Local-Sensitivity-mechanism-for-$p$} (\textsf{LSP}), a mechanism to estimate the value of $p$ under the protection of differential privacy. \textsf{LSP} exploits the properties of local sensitivity, which is basically the sensitivity of $p$ on adjacent datasets of the actual dataset $D$, to estimate $p$ with high precision. Then, we propose \emph{Two-Laplace-mechanism-for-$\lambda$ } (\textsf{TLL}), a mechanism to estimate the value of $\lambda$ under the protection of differential privacy. \textsf{TLL} uses the estimated value of $p$ from \textsf{LSP} and applies two Laplace mechanisms independently to estimate $\lambda$ with high precision. 

\vspace{2mm}The contributions of our work can be summarized as follows: 
\begin{itemize}
    \item We propose two private mechanisms,  \textsf{LSP} and \textsf{TLL},  for publishing the parameters $p$ and $\lambda$ of Weibull distribution for parametric survival analysis. We prove that the two mechanisms satisfy the definition of differential privacy.
    \item We show that the proposed mechanisms outperform other differentially private techniques in terms of median absolute error (MdAE) on four real datasets. In addition, our proposed mechanisms can produce meaningful output even at very low privacy budgets.
\end{itemize}

\vspace{2mm}Table~\ref{tab:notation} lists the notations along with their descriptions  used in this paper. The rest of the paper is organized as follows. In Section~\ref{sec:work}, we review the related work on differential privacy and privacy-preserving survival analysis.
In Section~\ref{sec:para}, we formally introduce the parametric model with Weibull distribution and derive the system of equations used to estimate Weibull's parameters.
In Section~\ref{sec:overview}, we present our proposed solution. In Section~\ref{sec:solution}, we present the proof of privacy protection. We discuss the experimental results in Section~\ref{sec:exp}. We conclude the paper in Section~\ref{sec:con}.

\section{Related work}
\label{sec:work}

Differential privacy \cite{dwork2004privacy} is a cryptography-based privacy framework which has been extensively studied in recent years. 
Dwork \emph{et al.} \cite{dwork2006calibrating} proposed the Laplace mechanism which calibrates noise to the global sensitivity of the output.
McSherry and Talwar \cite{mcsherry2007mechanism} proposed the exponential mechanism, which is commonly used as a general framework for many differentially private mechanisms \cite{NIPS2012_4548, gaboardi2013linear, zhang2015private}. 
Nissim \emph{et al.} \cite{nissim2007smooth} proposed smooth sensitivity which calibrates the noise to an upper bound of the local sensitivity of the dataset. Zhang \emph{et al.} \cite{zhang2015private} proposed to apply local sensitivity to the exponential mechanism for graph counting problems.

For survival analysis, O'Keefe \emph{et al.} \cite{o2012confidentialising} showed the possibility in leaking personal information from the outputs of survival analysis queries to database system, and then proposed many techniques in order to protect the privacy. For non-parametric models, they proposed to smooth the survival plot and add a small amount of noise to the result. For parametric models and semi-parametric models, they proposed to sample 95\% of the dataset with robust estimators, and then round the estimated results. In fact, these privacy-preserving techniques are based on the previous work from the field of statistical disclosure control \cite{cleveland1979robust, dandekar2004maximum, duncan1991microdata, doyle2001confidentiality}.
Besides, there are also other privacy-preservation techniques for specific models in survival analysis. 
Fung \emph{et al.} \cite{yu2008privacy, fung2008privacy} proposed to use a random linear kernel in Cox regression which is then applied to lung cancer survival analysis. Chen and Zhong \cite{chen2011privacy} proposed a privacy-preserving model for comparing survival curves using the log-rank test.

\begin{table}[!t]
	\centering
	\caption{Notation.}
	\renewcommand{\arraystretch}{1.2}	
	\begin{tabular}{ |c |  p{60mm}| }
		\hline			
		$d(D,D^\prime)$  & Distance between two datasets \\
		$t_i$  & Time until the event of interest occurs \\
		$d_i$  & Censoring indicator \\
		$S(t)$  & The survival function\\
		$h(t)$  & The hazard function\\
		$\text{Pr}[X]$  & The probability of $X$\\
		$\mathcal M$  & Differentially private mechanism \\
		$\epsilon$  & Privacy budget \\		
		$\mathcal S^{(k)}(D)$  & Boundary at distance $k$ computed on dataset $D$\\
		$\Delta_f$  & The sensitivity of $f$ \\		
		$\mathcal U(p, D)$  & The utility function on dataset $D$\\		
		$\text{Laplace}(s)$  & A zero-mean random variable with $pdf(x; s) = \frac 1 {2s}\exp\left(-|x|/s\right)$ \\		
		$K$  & Number of rungs in the ladder \\			
		$X \propto Y$  & X equals $Y$ times a constant \\			
		\hline  
				\end{tabular}
			\label{tab:notation}
\end{table}

\section{Parametric  Model}
\label{sec:para}
Let $T$ be a random variable representing the time until an event of interest occurs. Two characteristic functions of $T$ are the survival function and the hazard function.

\begin{definition}[Survival function \cite{kleinbaum1996survival}]
$S(t)$ is the probability of survival during the time interval $[0, t]$.
\begin{align}
 S(t) = \mbox{Pr}\left[ T > t \right].
 \end{align}

\end{definition}

\begin{definition}[Hazard function \cite{kleinbaum1996survival}]  
	The hazard function $h(t)$ is the risk of death at time $t$.
\begin{align}
h(t) = \lim_{\Delta t \to 0}{ \frac{\mbox{Pr}\left[t \leq T < t + \Delta t ~|~ T \geq t  \right]} {\Delta t}}.
\end{align} 

\end{definition} 

This work focuses on a parametric model which assumes that $T$ follows the Weibull distribution with the following cumulative distribution function: $$cdf(t; \lambda, p) = 1- \exp\left(- \left(t/ \lambda \right)^p\right) ~~~(t  \geq 0),$$ where $p>0$ is the shape parameter and $\lambda > 0$ is the scale parameter of the distribution. The Weibull distribution is used as the underlying distribution of $T$ due to its effectiveness in representing survival processes. In fact, it is currently the most common model used for parametric survival analysis. We have $S(t; \lambda, p) =  \exp\left(- \left(t/ \lambda \right)^p\right)$ and  $h(t; \lambda, p) =p t^{p-1}/{\lambda^p}$. We apply this model to a dataset of $n$ individuals, $D = \left\{ (t_i, d_i) \mid i = 1 \dots n \right\}$, where $t_i$ is the survival time of $i^{th}$ person and $d_i$ is the death indicator. $d_i = 1$ indicates that  $i^{th}$ person is dead at time $t_i$ and $d_i = 0$ indicates that $i^{th}$ person is alive at time $t_i$, and the exact survival time is censored (also known as \emph{right censoring}). Here, we also assume that for any $i=1\dots n$, $t_i$ is normalized to guarantee $t_i \in [\exp(-\omega), 1]$, where $\omega > 0$ is a hyper-parameter of our problem. We will revisit $\omega$ later as it is needed for our solution. To estimate the values of $p$ and $\lambda$, the traditional method is the maximum likelihood estimator (MLE) \cite{kleinbaum2012parametric} which maximizes the following log-likelihood function:
\begin{align*}
\ell (\lambda, p) &= \ln \prod_{i=1}^{n}  h(t_i; \lambda, p)^{d_i} \cdot S(t_i; \lambda, p)\\
& =\sum_{i=1}^n \left\{d_i \cdot \left( \ln p + (p-1) \cdot \ln t_i -p\ln \lambda  \right) - t_i^p / {\lambda^p} \right\}.
\end{align*}
We can maximize $\ell(\lambda, p)$  by setting the derivatives with respect to $\lambda$ and $p$ equal to zero. We derive the following system of equations:
\begin{numcases} {}
 \frac {\sum_i t_i^p \ln t_i}{\sum_i t_i^p} = \frac 1 p + \frac {\sum_i d_i \ln t_i} {\sum_i d_i},   \label{eqn:wei} \\
 ~~~~\lambda^p~~~~~~  = \frac {\sum_{i} t_i^p} {\sum_{i} d_i} .  \label{eqn:lambda} \label{eqn:lam}
\end{numcases}

For methods without privacy protection, we could use the Newton-Raphson method \cite{ypma1995historical} to compute $p$ as the root of Equation~\eqref{eqn:wei} and substitute that value into Equation~\eqref{eqn:lambda} to get  $\lambda$. However, as the values of $p$ and $\lambda$ are published and available to the public, we implicitly leak the information about the individuals in the dataset $D$ as the values of $p$ and $\lambda$ are derived from the values of $t_i$'s and $d_i$'s. Moreover, the MLE does not have a mechanism for controlling the amount of leaked information when $\lambda$ and $p$ are published. Our work aims to propose private mechanisms for estimating $p$ and $\lambda$ with high accuracy, which also allow us to control the amount of information to be leaked
  to the public.

\section{Our Solution }
\label{sec:overview}

In this section, we present our approach which consists of two steps. In the first step, we propose \emph{Local-Sensitivity-mechanism-for-$p$} (\textsf{LSP}), a private mechanism which allows us to estimate the value of $p$ as the root of Equation~\eqref{eqn:wei}. In the second step, we propose \emph{Two-Laplace-mechanism-for-$\lambda$ } (\textsf{TLL}), a private mechanism which uses the value of $p$ from \textsf{LSP} to estimate  $\lambda$ from Equation~\eqref{eqn:lambda}. For each step, we will spend $\epsilon/2$ privacy budget. By the composability property of differential privacy \cite{dwork2013algorithmic}, it is guaranteed that the pair ($\lambda, p$) to be published is $\epsilon-$differentially private.

\subsection{Local-Sensitivity Mechanism for Estimating  $p$ (\textsf{LSP})}

The proposed mechanism for estimating the shape parameter $p$, \textsf{LSP}, is based on our  experimental observation: for small sized datasets, the estimated value of $p$ from Equation~\eqref{eqn:wei} can fluctuate a lot when we switch from one dataset to its adjacent datasets. However, for large sized and medium sized datasets, the fluctuation is very little. This observation suggests that techniques which are based on additive noise calibrated with global sensitivity, i.e., Laplace mechanism, are not applicable. Instead, our proposed \textsf{LSP} mechanism calibrates the noise with local sensitivity of $p$. Local sensitivity can be understood as the fluctuation of the output when we 
switch from the \emph{actual dataset $D$} to its adjacent datasets. It is different from the global sensitivity which considers any pair of adjacent datasets. Local sensitivity is first proposed by Kobbi \emph{et al.} \cite{nissim2007smooth} and later used by Jun \emph{et al.} \cite{zhang2015private} on privately counting sub-graphs. Our proposed \textsf{LSP} mechanism is an extension of Jun \emph{et al.}'s idea from the discrete domain into the real domain.
\pgfplotsset{every axis/.append style={
                        },
    cmhplot/.style={mark=none,line width=1pt,<->},
    soldot/.style={only marks,mark=*, mark size=1.5pt},
    holdot/.style={fill=white,only marks,mark=*,mark size=1.5pt},
}

\tikzset{>=stealth}

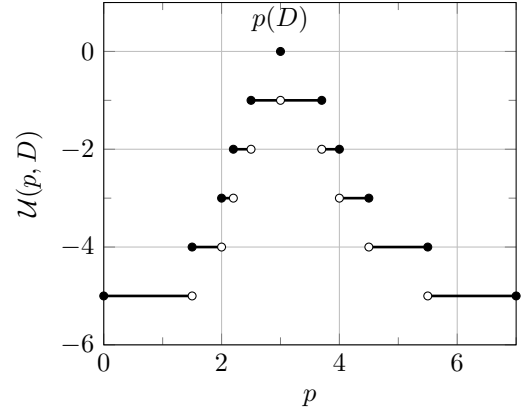
\begin{figure}[]
\hspace{5.5mm}\begin{tikzpicture}
\begin{axis}[scale=0.8, xmin=0,xmax=7, ymin=-6, ymax=1,grid=major, minor tick num =1, xlabel={$p$}, ylabel={$\mathcal U(p, D)$}]
        \addplot[cmhplot,-,domain=2.5:3]{-1};
        \addplot[cmhplot,-,domain=3:3.7]{-1};

        \addplot[cmhplot,-,domain=2.2:2.5]{-2};
        \addplot[cmhplot,-,domain=3.7:4]{-2};
        
        \addplot[cmhplot,-,domain=2.0:2.2]{-3};
        \addplot[cmhplot,-,domain=4.0:4.5]{-3};
        
        \addplot[cmhplot,-,domain=1.5:2.0]{-4};
        \addplot[cmhplot,-,domain=4.5:5.5]{-4};
        
        \addplot[cmhplot,-,domain=0.0:1.5]{-5};
        \addplot[cmhplot,-,domain=5.5:7.0]{-5};
        
                \addplot[soldot] coordinates{(3,0)} node[label={90:{$p(D)$}}]{};
        \addplot[holdot] coordinates{(3,-1)(2.5,-2)(3.7,-2)(2.2,-3)(4.0,-3)(2.0,-4)(4.5,-4)(1.5,-5)(5.5,-5)};
        \addplot[soldot] coordinates{(2.5,-1) (3.7,-1) (2.2,-2) (4.0,-2) (2,-3) (4.5,-3) (1.5,-4) (5.5,-4) (0,-5) (7,-5)};

\end{axis}
\end{tikzpicture}

  \caption{An example of the utility function $\mathcal U(p, D)$. The utility function has 6 rungs. The root of Equation~\eqref{eqn:wei} on dataset $D$ is $p(D)$, which is located at rung $0$. }
  \label {fig:ladder}
\end{figure} 
The \textsf{LSP} mechanism is constructed based on the exponential mechanism. We define a utility function $\mathcal U(p, D)$, which is referred to as the \emph{usefulness} of publishing the value $p$ with respect to the dataset ${D}$. Let $\Delta_{\mathcal U}$ be the sensitivity of the function $\mathcal U(p, D)$, and for any pair of adjacent datasets $D$ and $D^\prime$, we have:
\begin{align}
\Delta_{\mathcal U} = \max_{p, D, {D}^\prime: d(D, D^\prime) = 1} { \left| \mathcal U(p, {D}) - \mathcal U(p,{D}^\prime) \right| }.
\end{align} 
The output of the exponential mechanism is sampled from the following probability distribution:
\begin{align}
\mbox{Pr}\left[  \mathcal M({D}) = p \right] \propto \exp \left(  \frac{ \epsilon \cdot \mathcal U(p, D)}{2 \Delta_{\mathcal U}}  \right).
\end{align}
It was proved that the output sampled from the above distribution is $\epsilon-$differentially private (see \cite{mcsherry2007mechanism} for the detailed proof). 

In fact, \textsf{LSP} defines a utility function $\mathcal U(p, D)$ with the shape of a ladder as illustrated in Figure~\ref{fig:ladder}. Each rung of the ladder is defined as the set of points where the root of Equation~\eqref{eqn:wei} is placed when we replace the actual dataset $D$ by a hypothetical dataset $D^{(k)}$ such that $d(D, D^{(k)}) \leq -k$, where $k$ is the rung's level. For example, at  rung level $0$ we get the exact value of $p$ from Equation~\eqref{eqn:wei} on dataset $D$; while at rung level $-1$, the root of Equation~\eqref{eqn:wei} on any adjacent dataset of the dataset $D$ is placed in the segment bounded by two extremal points of $p$ at which  the values of the utility function $\mathcal U(p, D)$ are equal to $-1$.  However, it is very hard to compute exactly the extremal points at each rung of the utility function. Instead, we propose approximated intervals such that the interval at each rung will contain extremal points.  The mathematical definition of these intervals are as follows.

\begin{definition}[Local-Sensitivity Interval (LSI) at distance $k$]
	The interval $\mathcal S^{(k)}(D) = [l^{(k)}(D), u^{(k)}(D)]$ 
	is the local-sensitivity interval at distance $k$ of parameter $p$ if and only if for any dataset $D^{(k)}$ such that $d(D, D^{(k)}) \leq k$, then $p(D^{(k)})  \in \mathcal S^{(k)}(D)$, where $p(D^{(k)})$ denotes the root of Equation~\eqref{eqn:wei} on dataset $D^{(k)}$.
\end{definition}

By constructing the utility function $\mathcal U(p, D)$ as a ladder of rungs created by intervals $[l^{(k)}(D), u^{(k)}(D)]$, we will prove in Section~\ref{sec:solution} that the sensitivity of $\mathcal U(p, D)$  is equal to $1$. This will consequently prove that our mechanism is differentially private. 
However, we need a method to compute LSI first.

\subsubsection{Computing the LSI at distance $k$}

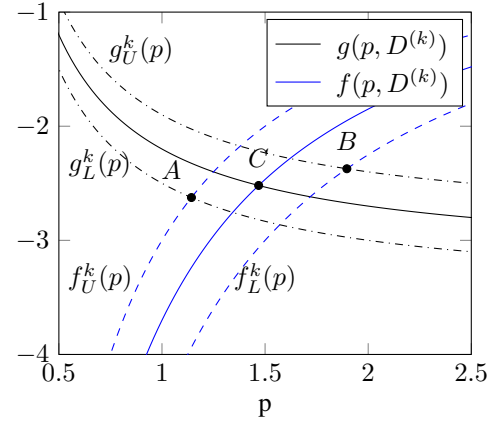
\begin{figure}[t]
  \centering
\begin{tikzpicture}
\begin{axis}[scale=0.8, xmin=0.5,xmax=2.5, ymin=-4, ymax=-1,
legend entries={ {} {$g(p,D^{(k)})$} , {} {$f(p, D^{(k)})$} }, xlabel={p} ]
\addplot [domain=0.02:4,samples=200,color=black] {1/x - 3.2} ;
\addplot [domain=0.02:4,samples=200,color=blue] {-3.7/x };

\addplot [domain=0.02:4,samples=200,color=black,dashdotted] {1/x - 3.5};
\addplot [domain=0.02:4,samples=200,color=black,dashdotted] {1/x - 2.9};
\addplot [domain=0.02:4,samples=200,color=blue,dashed] {-3/x  };
\addplot [domain=0.02:4,samples=200,color=blue,dashed] {-4.5/x };

\addplot[soldot] coordinates{(1.46875,-3.7/1.46875)} node[label={90:{$C$}}]{};
\addplot[soldot] coordinates{(1.89655,-4.5/1.89655)} node[label={90:{$B$}}]{};
\addplot[soldot] coordinates{(1.14286,-3/1.14286)} node[label={94:{$A$}}]{};

\node [label={[shift={(0.7,-2.7)}]$g_L^k(p)$}] {Node};
\node [label={[shift={(0.9,-1.7)}]$g_U^k(p)$}] {Node};

\node [label={[shift={(1.5,-3.7)}]$f_L^k(p)$}] {Node};
\node [label={[shift={(0.7,-3.7)}]$f_U^k(p)$}] {Node};

\addplot [domain=0.0:4,samples=200,black] {0};
\end{axis}
\end{tikzpicture}

  \caption{The estimated value of $p$ on dataset $D^{(k)}$ is determined by the intersection $C$ of two curves $f(p, D^{(k)})$ and $g(p, D^{(k)})$. The $p$-coordinate of point $C$ is bounded by the $p$-coordinate of point $A$ and the $p$-coordinate of point $B$ where point $A$ is the intersection between $f_U^k(p, D)$ and $g_L^k(p, D)$; and point B is the intersection between $f_L^k(p, D)$ and $g_U^k(p, D)$.
    }
  \label {fig:intersection}
\end{figure}

 For computing LSI, we introduce the boundary functions which are the upper bounds (resp., lower bounds) of the left hand side, LHS, (resp., right hand side, RHS) of Equation~\eqref{eqn:wei}. We pick these boundary functions to guarantee that the LHS and the RHS of Equation~\eqref{eqn:wei} are always bounded by the boundary functions when we apply Equation~\eqref{eqn:wei} to any hypothetical dataset $D^{(k)}$ at distance $k$ to the actual dataset $D$.  Consequently, this property will guarantee that the root of Equation~\eqref{eqn:wei} on  $D^{(k)}$ is also bounded by the intersections of these boundary functions. Let $f(p, D^{(k)})$ and $g(p, D^{(k)})$ be the LHS and the RHS of Equation~\eqref{eqn:wei} on dataset $D^{(k)}$, $e$ be the base of the natural logarithm, and $t_i^*$ be the $i^{th}$ smallest survival time in the dataset. We can bound $f(p, D^{(k)})$ and $g(p, D^{(k)})$ by four families of functions:
 \begin{align*}
f_L^k(p, D) &= \frac { \sum_i t^p_i \ln t_i   - \frac k {ep} }  {\sum_{i=1}^{n-k} (t_i^*)^p}, \\
f_U^k(p, D) &= \frac {  \sum_i t^p_i \ln t_i  + \frac k {ep} }  {\sum_{i=1}^{n} t_i^p+k}, \\
g_L^k(p, D) &=  \frac 1 p + \frac {\sum_i d_i \ln t_i- k\omega} { \sum_i d_i -k }, \\
g_U^k(p, D) &= \frac 1 p + \frac {\sum_i d_i \ln t_i+ k\omega} {\sum_i d_i +k}.
\end{align*}
The function $f_L^k(p, D)$ (resp., $f_U^k(p, D)$) is chosen to be the lower bound (resp., upper bound) of $f(p, D^{(k)})$. These functions are derived from the fact that the minimum value of the function $h(t) = t^p \ln t$ for a fixed $p$ and $t \in [\exp(-\omega), 1]$ is $-\frac{1}{ep}$. The function $g_L^k(p, D)$ (resp., $g_U^k(p, D)$) is the lower bound  (resp., upper bound) of $g(p, D^{(k)})$. This is from the fact that the minimum value of $k(t) = \ln t$  for $t \in [\exp(-\omega), 1]$ is $-\omega$. Our observation is that the estimation of $p$ on $D^{(k)},$ referred as $p_C,$ is bounded by an interval $[p_A,  p_B]$, where  $p_A$ is the root of equation $f_U^k(p, D) = g_L^k(p, D)$ and $p_B$ is the root of equation  $f_L^k(p, D) = g_U^k(p, D)$. This observation is illustrated in Figure~\ref{fig:intersection}.

\begin{algorithm}[]
	\caption{{Compute LSIs}}\label{mech:weibullls}
	\begin{algorithmic}[1]
		\Statex \textbf{Input:} Dataset D, number of rungs K
		\Statex \textbf{Output:} lower bounds $l$, upper bounds $u$
		\Statex \textbf{Requirement:} $0\leq p \leq \gamma$
						\State $l^{(0)} = u^{(0)} = $ root of Equation~\eqref{eqn:wei}
		\For{ $i = 1 \dots K$}
		\State $l^{(i)} = $ root of equation $f^k_U(p) = g^k_L(p)$
		\State $u^{(i)} = $ root of equation $f^k_L(p) = g^k_U(p)$
		\EndFor
		\State $l^{(K+1)} = 0$,  $u^{(K+1)} = \gamma$ 
		\State \Return $l, u$
			\end{algorithmic}
\end{algorithm}

Algorithm~\ref{mech:weibullls} computes the lower bounds and upper bounds of $p$. At Line 1, we compute the exact solution of Equation~\eqref{eqn:wei} on dataset $D$ and use this value as the first lower bound and upper bound. The \textbf{for} loop at Lines 2-5 gives the $K$ rungs defined by the pairs of lower bounds $l^{(i)}$ and upper bounds $u^{(i)}$, where $K$ is a hyper-parameter of the solution which will be decided based on the trade-off between accuracy and runtime.  In detail, we use the Newton-Raphson method to compute the roots of equations at Lines 3-4. The \emph{floor rung} is given at Line 6 to guarantee that all values of $p$ are contained in the interval $[0, \gamma]$,  where the hyper-parameter $\gamma$ will be decided based on the nature of the survival analysis problem.
Line 7 returns all the lower bounds and upper bounds of LSIs in two lists $l$ and $u$.

\subsubsection{Publishing the value of $p$ by exponential mechanism}

\textsf{LSP} uses the exponential mechanism as the framework to estimate the value of $p$. We define the following utility function which is based on LSI $\mathcal S^{(k)}(D)$ computed from the previous step.

\begin{definition}[Ladder-shaped  utility function]
	The ladder-shaped utility function $\mathcal U(p, D)$ is defined as: 
	\begin{align} 
	\mathcal U(p, D) = -k \mbox{ iff } p \in &
	\left[l^{(k)}(D),~l^{(k-1)}(D)\right)  \cup  \nonumber\\
	&\left(u^{(k-1)}(D),~u^{(k)}(D)\right],
		\end{align} 		where $k = 1 \dots (K+1).$

\end{definition}

From the definition of the utility function, we propose the \textsf{LSP} mechanism (Algorithm~\ref{mech:ls}) for estimating the value of $p$.

\begin{algorithm}[]
	\caption{{Local-Sensitivity Mechanism for $p$ (\textsf{LSP})}}\label{mech:ls}
	\begin{algorithmic}[1]
		\Statex \textbf{Input:} Dataset $D$, number of rungs K, privacy budget $\epsilon/2$, lower bounds $l$, upper bounds $u$
		\Statex \textbf{Output:} $p$
						\For{ $i = 1 \dots K+1$}
		\State ${level}^{(i)} =  \left[l^{(i)},l^{(i-1)}\right)  \cup \left(u^{(i-1)},u^{(i)}\right] $
		\State $length^{(i)} =  u^{(i)} - u^{(i-1)} + l^{(i-1)} - l^{(i)}$
		\State $weight^{(i)} = length^{(i)} \cdot  \exp \left(-i \frac{\epsilon}{4} \right)$
				\EndFor
								\State Sampling a level from distribution: $\mbox{Pr} [ {level}^{(i)} ] = \frac {{weight}^{(i)}} { {sum}({weight})}$
		\State \Return a value of $p$ sampling from uniform distribution over the chosen level at Line 6
	\end{algorithmic}
\end{algorithm}

Algorithm~\ref{mech:ls} 
draws a randomized value of $p$ from a probability density function which is constructed from the ladder-shaped utility function. Line 2 in Algorithm~\ref{mech:ls} specifies $level^{(i)}$ as the union of two intervals in the  $i^{th}$ rung of the utility function. The values of $p$ in $level^{(i)}$ share the same probability density $\exp(-i \epsilon/4)$. The probability at level $i$ is accumulated in $weight^{(i)}$ at Line 4. Line 6 draws a random rung from its weights and Line 7 draws a randomly uniform value of $p$ from the chosen rung at Line 6. In Section~\ref{sec:solution}, we will prove that the \textsf{LSP} mechanism is $\epsilon/2-$differentially private.

\subsection{Two-Laplace Mechanism for Estimating $\lambda$ (\textsf{TLL})}

In order to estimate the scale parameter $\lambda$, we propose \emph{Two-laplace-mechanism-for-$\lambda$} (\textsf{TLL}), which is based on the Laplace mechanism proposed by Dwork \emph{et al.} \cite{dwork2006calibrating}.  Laplace mechanism is a simple mechanism for achieving differential privacy by calibrating the additive noise to the global sensitivity of the result. Let's assume that we want to publish the value of a function $f(\cdot)$ which takes the dataset as the input and returns a real value. The \emph{global sensitivity} of function $f(\cdot)$ is 
$$	\Delta_f = \max_{{D}, {D}^\prime: d(D, D^\prime) = 1} { | f({D}) - f({D}^\prime)  | },$$ where $(D, D^\prime)$ is any pair of adjacent datasets.
Laplace mechanism $\mathcal M(D)$ returns a randomized version of $f(D)$ by adding Laplacian noise, $
\mathcal M({D}) = f({D})   +  \text{Laplace}\left(\Delta_f/\epsilon\right)
,$ where $\text{Laplace}\left({\Delta_f}/{\epsilon}\right)$ is a random Laplacian variable
  It was proved that $\mathcal M$ is $\epsilon-$differentially private (see \cite{dwork2006calibrating} for the proof).

 We observe that for small datasets, the value of $\lambda = \sqrt[p]{\frac{\sum_i t_i^{p}}{\sum_i d_i}}$ can vary a lot between adjacent datasets. Therefore, we cannot apply the Laplace mechanism to $\lambda$ directly because its global sensitivity is too large. Since the global sensitivities of  $\sum_i d_i$ and $\sum_i t_i^{p}$  are equal to $1$, instead of adding noise to $\lambda$ directly, our proposed \textsf{TLL} mechanism spends $\epsilon/4$ privacy budget to publish $\sum_i d_i$ and another $\epsilon/4$ privacy budget to publish $\sum_i t_i^{p}$. By the composability property of differential privacy, \textsf{TLL} is $\epsilon/2$-differentially private. The formal proof is given in Section~\ref{sec:solution}.

\begin{algorithm}[]
	\caption{{Two-Laplace Mechanism for $\lambda$ (\textsf{TLL})}}\label{mech:plam}
	\begin{algorithmic}[1]
		\Statex \textbf{Input:} Dataset ${D}$, privacy budget $\epsilon/2$, $p$
		\Statex \textbf{Output:} $\lambda$
		\State Compute $\delta = \sum _ { i = 1} ^ n  d_i  + \text{Laplace}(4/\epsilon)$
		\State Compute $\tau = \sum _{i=1} ^n t_i^p   + \text{Laplace}(4/\epsilon)$
		\State Get $\lambda = \sqrt[p] {{\tau}/ {\delta} }$
	\end{algorithmic}
\end{algorithm}

Algorithm~\ref{mech:plam} gives the proposed \textsf{TLL} mechanism. Lines 1-2 in Algorithm~\ref{mech:plam} compute the two quantities $\delta$ and $\tau$ by adding  independent Laplacian noises to $\sum_i d_i$ and  $\sum_i t_i^{p}.$ 
 Line 3 in Algorithm~\ref{mech:plam} returns the value of $\lambda$ as derived from Equation~\eqref{eqn:lam}.

\section{ Proof of Privacy Protection}
\label{sec:solution}

\subsection{Proof for the Proposed \textsf{LSP} Mechanism}

In order to prove the differentially private protection of the \textsf{LSP} mechanism, we need to introduce a constraint on the LSI $\mathcal S^{(k)}(D)$. We will use this constraint later to prove that the sensitivity of the utility function is equal to 1.

\begin{definition}[Ladder constraint] Local-sensitivity intervals $\mathcal S^{(k)}(D)$ are  \textbf{ladder local-sensitivity intervals} if and only if $\mathcal S^{(k-1)}(D) \subset \mathcal S^{(k)}(D)$ and for any pair of adjacent datasets $D$ and $D^\prime$, $\mathcal S^{(k)}(D) \subset \mathcal S^{(k+1)}(D^\prime)$.
\end{definition}

We now prove that the output of the first step in our solution satisfies the ladder constraint.

\begin{lemma} For any pair of adjacent datasets $D$ and $D^\prime$, we have: \\ 
	(a)~~ $g_L^{k+1}(p, D^\prime) \leq g_L^k(p, D)$ and  $g_U^k(p, D) \leq g_U^{k+1}(p, D^\prime)$,  and\\
	(b)~~ $f_L^k(p, D) \geq f_L^{k+1}(p, D^\prime)$ and $f_U^k(p, D) \leq f_U^{k+1}(p,D^\prime)$.\label{lm:main}
\end{lemma}
\begin{IEEEproof} 
	(a) We have $g_L^k(p, D) - g_L^{k+1}(p, D^\prime)  \geq 0$ and  $g_U^k(p, D) - g_U^{k+1}(p, D^\prime) \leq 0$. Therefore, $g_L^{k+1}(p, D^\prime) \leq g_L^k(p, D)$ and  $g_U^k(p, D) \leq g_U^{k+1}(p, D^\prime)$.

(b) We have $f_L^k(p, D) - f_L^{k+1}(p, D^\prime) \geq 0$ and $f_U^k(p, D) - f_U^{k+1}(p,   D^\prime) \leq 0$. Therefore, $f_L^k(p, D) \geq f_L^{k+1}(p, D^\prime)$ and $f_U^k(p, D) \leq f_U^{k+1}(p, D^\prime)$.
\end{IEEEproof}

\begin{table*}[!pt]
	
	\renewcommand{\arraystretch}{1.3}	
	\caption{Datasets used in the experiments.}
	\centering
	\begin{tabular}{ |c| | c|c | p{100mm}| }
		\hline
		\thead{	Dataset } & \thead{Size} & \thead{\#uncensored} & \thead{Description} \\ \hline\hline
		\textsf{FL} & $7874$ & $2169$ & The dataset on the relationship between serum free light chain (FLC) and mortality. \\		\hline	
		\textsf{TB} & $16116$ & $1761$ & The Medical Birth Registry of Norway on the time between second and third births.\\			 \hline
		\textsf{WT} & $21685$ & $18615$ & The dataset on the unemployment time of people in Germany.\\		\hline	
		\textsf{SB} & $53558$ & $16341$ & The Medical Birth Registry of Norway on the time between the first and second births.\\			
		\hline  
	\end{tabular}
	
	\label{tab:dataset}
\end{table*}

\begin{theorem} Let $\mathcal S^k (D) = [l^{(k)}(D),u^{(k)}(D)]$, then  \\~~~~~(a)~~~ $\mathcal S^{(k-1)} (D) \subset \mathcal S^{(k)}(D)$, and \\~~~~~(b)~~~ $\mathcal S^{(k)} (D) \subset \mathcal S^{k+1} (D^\prime)$.\label{lm:ladderconstraint}
\end{theorem}
\begin{IEEEproof} 
		(a) We have $f^{k-1}_U (p,D) \leq f^K_U (p,D)$ and $g^{k-1}_L (p,D) \geq g^k_L(p, D)$,  and  $g_L^{k-1}(p, D)$ and $g_L^k(p, D)$ are non-increasing functions. Thus, we have $l^{(k-1)}(D) \geq l^{(k)} (D)$.
	We also have $f^{k-1}_L (p,D) \leq f^k_L (p,D)$ and $g^{k-1}_U (p,D) \geq g^k_U (p, D)$, and  $g_U^{k-1}(p, D)$ and $g_U^k(p, D)$ are non-increasing functions. Thus, we have $u^{(k-1)}(D) \leq u^{(k)} (D)$.
Therefore, $\mathcal S^{k-1}(D) \subset \mathcal S^{k}(D)$.

 (b) By Lemma~\ref{lm:main} and similar arguments as (a), we  have $l^{(k)}(D) \geq l^{(k+1)} (D^\prime)$ and  $u^{(k)}(D) \leq u^{(k+1)} (D^\prime)$.
Therefore, $\mathcal S^{k}(D) \subset \mathcal S^{k+1}(D^\prime)$.
\end{IEEEproof}

\begin{lemma} If $p \in \mathcal S^{(k)}(D)$, then $ - \mathcal U(p, D) \leq k$. 
	\label{lm:sub}
\end{lemma}
\begin{IEEEproof}By the definition of $\mathcal U(\cdot)$, if $-\mathcal U(p, D) = h$, then $p \in \mathcal S^{(h)}(D)$ and $p \notin \mathcal S^{(h-1)}(D)$. By the ladder constraint, we have $\mathcal S^{(i)}(D) \subset \mathcal S^{(h-1)}(D)$ for any $i < h - 1$. Therefore, if $p \notin \mathcal S^{(h-1)}(D)$, then $p \notin \mathcal S^{(i)}(D)$ for any $i \leq h - 1$. In other words, if $-\mathcal U(p, D) = h$, then $p \notin \mathcal S^{(i)}(D)$ for any $i \leq h - 1$.
	Now assuming $h > k$, from the above argument it implies that $p \notin \mathcal S^{(k)}$. This contradicts with the assumption of the lemma. Therefore, the lemma is proven by contradiction.
\end{IEEEproof}

\begin{lemma} The sensitivity of utility function $\mathcal U(p, D)$ is $1$.$$
	\max_{p, D, D^\prime: d(D, D^\prime) =1} | \mathcal U(p, D) - \mathcal U(p, D^\prime) | \leq 1.
	$$\label{lm:sen}\end{lemma}
\vspace{-4mm}\begin{IEEEproof}For simplicity, let $k$ be the short form for $-\mathcal U(p, D)$ and $h$ be the short form for $-\mathcal U(p, D^\prime)$. To prove the lemma, we need to prove that $k - 1\leq h \leq k+1$.
	Assuming that $h < k - 1$. We have $p \in \mathcal S^{(h)}(D^\prime)$. By the ladder constraint, we have $\mathcal S^{(h)}(D^\prime) \subset  \mathcal S^{(h+1)}(D) \subset \mathcal S^{(k-1)}(D)$. So $p \in \mathcal S^{(k-1)}(D)$. Lemma~\ref{lm:sub} implies $-\mathcal U(p, D) \leq k-1$ and this contradicts with the assumption that $-\mathcal U(p, D) = k$. Hence, $h \geq k + 1$.
	By the ladder constraint, we also have $p \in \mathcal S^{(k)}(D) \subset  \mathcal S^{(k+1)}(D^\prime)$. Finally, from Lemma~\ref{lm:sub}, we have $h \leq k + 1$. Therefore, the lemma is proven.\end{IEEEproof}

\begin{theorem} \textsf{LSP} is $ \epsilon/2-$differentially private.\end{theorem}

\begin{IEEEproof} Lemma~\ref{lm:sen} claims that the sensitivity of the utility function $\mathcal U(\cdot)$ is equal to $1$. Hence,  \textsf{LSP}  is $\epsilon/2-$differentially private by the property of the exponential mechanism.
\end{IEEEproof}

\subsection{Proof for the Proposed \textsf{TLL} Mechanism}

\begin{theorem} \textsf{TLL} is $ \epsilon/2-$differentially private.\end{theorem}

\begin{IEEEproof}  The global sensitivity of $\sum_i d_i$ is equal to $1$ because $d_i \in \{0, 1\}$. Therefore, the value $\delta$ in Algorithm~\ref{mech:plam} is $\epsilon/4-$differentially private. We have $t_i \in [0, 1]$, hence $t_i^p \in [0, 1]$ for $p > 0$. Therefore, the global sensitivity of $\sum_i t_i^p$ is equal to $1$. Consequently, the value $\tau$ in Algorithm~\ref{mech:plam} is $\epsilon/4-$differentially private. By the composability property of differential privacy, $\lambda = \sqrt[p]{\tau / \delta}$ is $ \epsilon/2-$differentially private. \end{IEEEproof}

\section{Performance Evaluation}
\label{sec:exp}

\subsection{Experiment Setup}

\header{\it Datasets.} We use four real datasets in the experiments. Table~\ref{tab:dataset} gives a summary of these datasets. 
\begin{itemize}
	\item 
\textbf{The FLchain dataset}  (\textsf{FL}) - It is obtained from a study on the association of the serum free light chain with higher death rates \cite{kyle2006prevalence, dispenzieri2012use}. The survival time of a patient is measured in days from enrollment until death. The censored cases are patients who are still alive at the last contact. 

\item \textbf{The time-to-second-birth} (\textsf{SB)}  and \textbf{time-to-third-birth} (\textsf{TB}) \textbf{datasets} - They are obtained from The Medical Birth Registry of Norway \cite{irgens2000medical}. The survival time is the time between the first and second births,  and between the second and third births respectively. The censored cases are women who do not have the second birth, and the third birth respectively, at the time the data are collected. 

\item \textbf{The Wichert dataset}  (\textsf{WT}) - It contains records on unemployment duration of people in Germany \cite{wichert2008simple}. The survival time is the duration of unemployment until having a  job again. The censored cases are the ones who do not have a new job at the time the data are collected. 
\end{itemize}

\begin{figure*}[h!]
	\centering 
	\includegraphics[width=\textwidth]{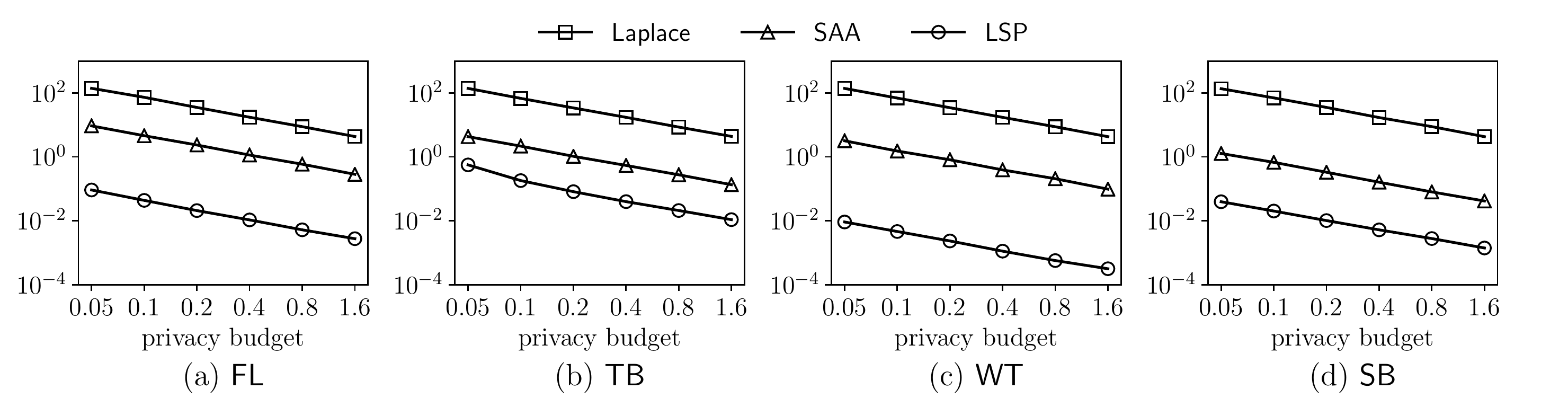}
	\caption{The median absolute errors (MdAE) on estimating the value of the shape parameter $p$.}
	\label{fig:p}
\end{figure*}

\begin{figure*}[h]
	\centering 
	\includegraphics[width=\textwidth]{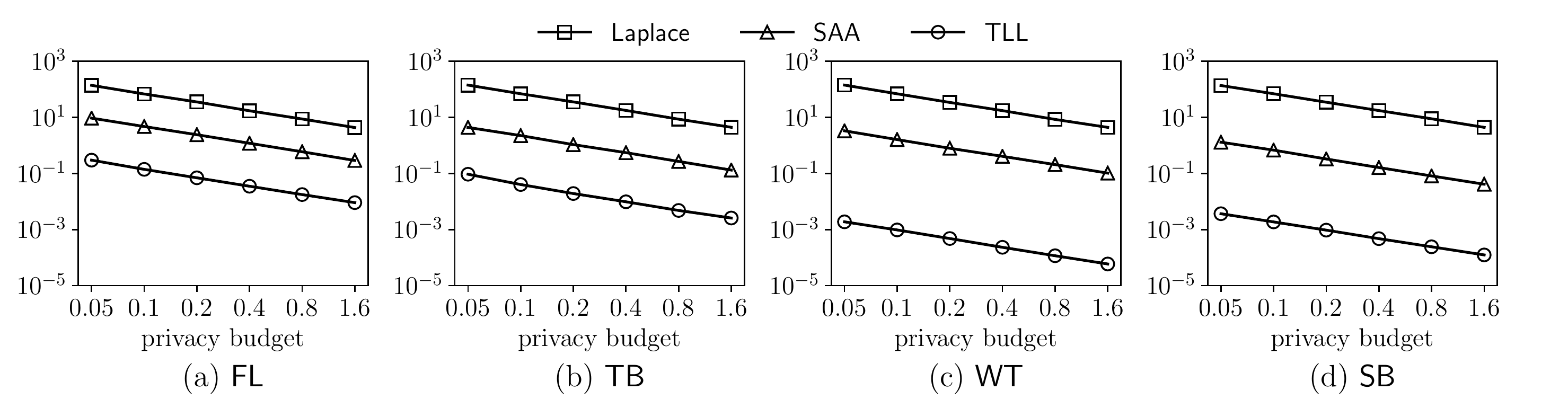}
	\caption{The median absolute errors (MdAE) on estimating the value of the scale parameter $\lambda$.}
	\label{fig:l}
\end{figure*}
\header{\it Evaluation metric.} We use the median absolute error (MdAE) for performance measurement. MdAE is defined as: \begin{align}\label{eqn:mre}
    \operatorname{MdAE} = median\left( \left\{  | x_i - x^* |, ~  i = 1 \dots t \right\} \right),
\end{align} where $x_i$ is an answer from private mechanism, $x^*$ is the exact value obtained from the traditional MLE approach without privacy protection and $t$ is the number of tries. We use MdAE instead of the commonly used mean squared error (MSE) because MdAE is more robust to outliers while MSE is sensitive to outliers.

\header{\it Baselines.} We compare our proposed mechanisms \textsf{LSP} and \textsf{TLL} with the Laplace mechanism and the \emph{sample-and-aggregate} (\textsf{SAA}) mechanism \cite{smith2008efficient,dwork2010differential}. 

\begin{itemize}
	\item The \textsf{Laplace} mechanism - It  adds Laplacian noise to the estimated values of $p$ and $\lambda$ independently. Each of them has the global sensitivity of $\gamma$ ($0 \le \lambda, p \leq \gamma$).
	\item The \textsf{SAA} mechanism - It divides the dataset into partitions of random non-overlap subsets, and uses MLE approach to estimate the parameters on these subsets. The estimated parameters are then aggregated together to derive the estimation of the whole dataset. The aggregated results are published by applying the Laplace mechanism. See \cite{smith2008efficient} for the details of the \textsf{SAA} mechanism.

\end{itemize}

\header{\it Hyper-parameter settings}. All experiments are repeated with $t=500$ times for statistical stability. The privacy budget consumed by each mechanism varies from  $0.05$ to $1.6$ for evaluating the accuracy of the results with different privacy budgets. Consequently, the total privacy budget $\epsilon$ consumed by private mechanisms to publish both $p$ and $\lambda$ varies from $0.1$ to $3.2$. For the \textsf{LSP} mechanism, we set the number of rungs $K=500$. For the  \textsf{SAA} mechanism, we set the 
number of subsets such that each subset has $500$ records on average. We set the upper bound $\gamma = 10$ for the values of $p$ and $\lambda$. Moreover, we also set $\omega = 6$. As such, we normalize all the survival times into the interval $[\exp(-6), 1].$

  \subsection{Experimental Results}

\subsubsection{Performance on estimating the shape parameter $p$}

 Figure~\ref{fig:p} shows the performance results of the Laplace mechanism, the \textsf{SAA} mechanism and our proposed \textsf{LSP} mechanism on estimating the parameter $p$. 

For the dataset \textsf{FL}, \textsf{LSP} significantly outperforms the \textsf{Laplace} and \textsf{SAA} mechanisms at all levels of privacy budgets. The MdAE of \textsf{LSP} is 100 times smaller than that of \textsf{SAA} and about 1500 times smaller than that of \textsf{Laplace}.
Even at the smallest privacy budget $0.05$, the MdAE of \textsf{LSP} is only $0.1$. In other words, at privacy budget $0.05$ we should expect the output of \textsf{LSP} is about 0.1 away from the exact output from the MLE. 
For the dataset \textsf{TB}, \textsf{LSP} still outperforms the \textsf{Laplace} and \textsf{SAA} mechanisms at all levels of privacy budgets. The MdAE of \textsf{LSP} is 7 to 13 times smaller than that of \textsf{SAA}. Even though the size of \textsf{TB} is 2 times bigger than the size of \textsf{FL}, the number of uncensored records of \textsf{TB} is smaller than the number of uncensored records of \textsf{FL}. Due to the smaller number of uncensored records, the estimated value of $p$  tends to be more sensitive to the change in the dataset. This causes the decrease in performance on \textsf{TB} when compared with \textsf{FL}.
For the dataset \textsf{WT}, our proposed \textsf{LSP} mechanism again significantly outperforms the \textsf{Laplace} and \textsf{SAA} mechanisms at all levels of privacy budgets. The MdAE of \textsf{LSP} is  300 times smaller than that of \textsf{SAA}. For the dataset \textsf{SB},  \textsf{LSP} still outperforms the \textsf{Laplace} and \textsf{SAA} mechanisms. Even though \textsf{SB}  is  bigger in size, it  has a smaller number of uncensored records than \textsf{WT}. Therefore,  the performance on \textsf{SB} is not as good as the performance on \textsf{WT}.

Overall, the \textsf{Laplace} mechanism performs quite consistently throughout the four datasets. The \textsf{SAA} mechanism tends to have better performance on bigger datasets.  \textsf{SB} has the size about 6 times bigger than \textsf{FL}. Meanwhile, the performance of \textsf{SAA} on \textsf{SB} is also about 6 times better than the performance on \textsf{FL}.  \textsf{LSP} has the best performance with useful output in all datasets, even at very small privacy budget, i.e., $0.05$. However, the performance of the \textsf{LSP} mechanism depends on the number of uncensored records more than the actual size of the datasets.

\subsubsection{Performance on estimating the scale parameter $\lambda$} 

Figure~\ref{fig:l} shows the performance results of the \textsf{Laplace} mechanism, the \textsf{SAA} mechanism and our proposed \textsf{TLL} mechanism on estimating the parameter $\lambda$. 

For the dataset \textsf{FL}, \textsf{TLL} outperforms the \textsf{Laplace} and \textsf{SAA} mechanisms. 
The MdAE of \textsf{TLL} is about 30 times smaller than that of SAA and about 450 times smaller than that of Laplace.  At the lowest privacy budget $0.05$, the MdAE of \textsf{TLL} is $0.297$ which is still acceptable because the exact value of $\lambda$ on this dataset is $2.6098$. In other words, the relative error in this case is about 11\%.  For the dataset \textsf{TB}, \textsf{TLL} still outperforms the \textsf{Laplace}  and \textsf{SAA} mechanisms. The performance of \textsf{TLL} on \textsf{TB} is about 3 times better than its performance on \textsf{FL}. For the dataset \textsf{WT}, \textsf{TLL} significantly outperforms the  \textsf{Laplace}  and \textsf{SAA} mechanisms. The performance of \textsf{TLL} is about 3 orders of magnitude better than that of the \textsf{SAA} mechanism and more than 4 orders of magnitude better than that of the Laplace mechanism. The increase in accuracy on this dataset is mainly due to the very accurate estimated value of $p$ from \textsf{LSP} and the bigger dataset size. These enable \textsf{TTL} to estimate $\lambda$ with very high accuracy.
For the dataset \textsf{SB}, \textsf{TLL} outperforms the other two mechanisms even though its performance is not as good as the performance on \textsf{WT}. The MdAE of \textsf{TLL} is about 300 times smaller than that of the \textsf{SAA} mechanism.

In summary, the \textsf{Laplace} mechanism has consistent performance throughout the four datasets. The \textsf{SAA} mechanism has improved performance for bigger datasets. Our proposed \textsf{TLL} mechanism outperforms  both \textsf{SAA} and \textsf{Laplace} mechanisms on all datasets and privacy budgets. Moreover, \textsf{TLL}  still outputs meaningful answers even at small privacy budgets.

\section{Conclusion and Future Work}
\label{sec:con}
In this paper, we have proposed differential private mechanisms for parametric survival analysis with Weibull distribution.  Our mechanisms exploit the property of local sensitivity to publish answers with high accuracy. We prove that our mechanisms guarantee privacy protection under differential privacy. The experiments on real datasets show that the performance of our mechanisms are better than the current state-of-the-art techniques and able to publish meaningful answers. This work is only the first step to explore the applications of differential privacy to protect privacy in survival analysis. For further work, we intend to investigate new mechanisms for parametric survival analysis with other probability distributions such as the log-normal distribution, and explore the possibilities of protecting Cox regression under differential privacy.
\balance

\bibliographystyle{./IEEEtran}
\bibliography{./IEEEabrv,./bare_conf}

\begin{thebibliography}{10}
\providecommand{\url}[1]{#1}
\csname url@samestyle\endcsname
\providecommand{\newblock}{\relax}
\providecommand{\bibinfo}[2]{#2}
\providecommand{\BIBentrySTDinterwordspacing}{\spaceskip=0pt\relax}
\providecommand{\BIBentryALTinterwordstretchfactor}{4}
\providecommand{\BIBentryALTinterwordspacing}{\spaceskip=\fontdimen2\font plus
\BIBentryALTinterwordstretchfactor\fontdimen3\font minus
  \fontdimen4\font\relax}
\providecommand{\BIBforeignlanguage}[2]{{%
\expandafter\ifx\csname l@#1\endcsname\relax
\typeout{** WARNING: IEEEtran.bst: No hyphenation pattern has been}%
\typeout{** loaded for the language `#1'. Using the pattern for}%
\typeout{** the default language instead.}%
\else
\language=\csname l@#1\endcsname
\fi
#2}}
\providecommand{\BIBdecl}{\relax}
\BIBdecl

\bibitem{miller2011survival}
R.~G. Miller~Jr, \emph{Survival analysis}.\hskip 1em plus 0.5em minus
  0.4em\relax John Wiley \& Sons, 2011, vol.~66.

\bibitem{klein2005survival}
J.~P. Klein and M.~L. Moeschberger, \emph{Survival analysis: techniques for
  censored and truncated data}.\hskip 1em plus 0.5em minus 0.4em\relax Springer
  Science \& Business Media, 2005.

\bibitem{kantoff2010overall}
P.~W. Kantoff, T.~J. Schuetz, B.~A. Blumenstein, L.~M. Glode, D.~L. Bilhartz,
  M.~Wyand, K.~Manson, D.~L. Panicali, R.~Laus, J.~Schlom \emph{et~al.},
  ``Overall survival analysis of a phase ii randomized controlled trial of a
  poxviral-based psa-targeted immunotherapy in metastatic castration-resistant
  prostate cancer,'' \emph{Journal of Clinical Oncology}, vol.~28, no.~7, pp.
  1099--1105, 2010.

\bibitem{fleming2000survival}
T.~R. Fleming and D.~Lin, ``Survival analysis in clinical trials: past
  developments and future directions,'' \emph{Biometrics}, vol.~56, no.~4, pp.
  971--983, 2000.

\bibitem{marubini2004analysing}
E.~Marubini and M.~G. Valsecchi, \emph{Analysing survival data from clinical
  trials and observational studies}.\hskip 1em plus 0.5em minus 0.4em\relax
  John Wiley \& Sons, 2004, vol.~15.

\bibitem{o2012confidentialising}
C.~O'Keefe, R.~S. Sparks, D.~McAullay, and B.~Loong, ``Confidentialising
  survival analysis output in a remote data access system,'' \emph{Journal of
  Privacy and Confidentiality}, vol.~4, no.~1, pp. 127--154, 2012.

\bibitem{dwork2006calibrating}
C.~Dwork, F.~McSherry, K.~Nissim, and A.~Smith, ``Calibrating noise to
  sensitivity in private data analysis,'' in \emph{Theory of
  cryptography}.\hskip 1em plus 0.5em minus 0.4em\relax Springer, 2006, pp.
  265--284.

\bibitem{dwork2004privacy}
C.~Dwork and K.~Nissim, ``Privacy-preserving datamining on vertically
  partitioned databases,'' in \emph{Advances in Cryptology--CRYPTO 2004}.\hskip
  1em plus 0.5em minus 0.4em\relax Springer, 2004, pp. 528--544.

\bibitem{mcsherry2007mechanism}
F.~McSherry and K.~Talwar, ``Mechanism design via differential privacy,'' in
  \emph{Foundations of Computer Science, 2007. FOCS'07. 48th Annual IEEE
  Symposium on}.\hskip 1em plus 0.5em minus 0.4em\relax IEEE, 2007, pp.
  94--103.

\bibitem{NIPS2012_4548}
M.~Hardt, K.~Ligett, and F.~Mcsherry, ``A simple and practical algorithm for
  differentially private data release,'' in \emph{Advances in Neural
  Information Processing Systems 25}, F.~Pereira, C.~Burges, L.~Bottou, and
  K.~Weinberger, Eds.\hskip 1em plus 0.5em minus 0.4em\relax Curran Associates,
  Inc., 2012, pp. 2339--2347.

\bibitem{gaboardi2013linear}
M.~Gaboardi, A.~Haeberlen, J.~Hsu, A.~Narayan, and B.~C. Pierce, ``Linear
  dependent types for differential privacy,'' in \emph{ACM SIGPLAN Notices},
  vol.~48, no.~1.\hskip 1em plus 0.5em minus 0.4em\relax ACM, 2013, pp.
  357--370.

\bibitem{zhang2015private}
J.~Zhang, G.~Cormode, C.~M. Procopiuc, D.~Srivastava, and X.~Xiao, ``Private
  release of graph statistics using ladder functions,'' in \emph{Proceedings of
  the 2015 ACM SIGMOD International Conference on Management of Data}.\hskip
  1em plus 0.5em minus 0.4em\relax ACM, 2015, pp. 731--745.

\bibitem{nissim2007smooth}
K.~Nissim, S.~Raskhodnikova, and A.~Smith, ``Smooth sensitivity and sampling in
  private data analysis,'' in \emph{Proceedings of the thirty-ninth annual ACM
  symposium on Theory of computing}.\hskip 1em plus 0.5em minus 0.4em\relax
  ACM, 2007, pp. 75--84.

\bibitem{cleveland1979robust}
W.~S. Cleveland, ``Robust locally weighted regression and smoothing
  scatterplots,'' \emph{Journal of the American statistical association},
  vol.~74, no. 368, pp. 829--836, 1979.

\bibitem{dandekar2004maximum}
R.~A. Dandekar, ``Maximum utility-minimum information loss table server design
  for statistical disclosure control of tabular data,'' in \emph{International
  Workshop on Privacy in Statistical Databases}.\hskip 1em plus 0.5em minus
  0.4em\relax Springer, 2004, pp. 121--135.

\bibitem{duncan1991microdata}
G.~T. Duncan and S.~Mukherjee, ``Microdata disclosure limitation in statistical
  databases: query size and random sample query control,'' in \emph{Research in
  Security and Privacy, 1991. Proceedings., 1991 IEEE Computer Society
  Symposium on}.\hskip 1em plus 0.5em minus 0.4em\relax IEEE, 1991, pp.
  278--287.

\bibitem{doyle2001confidentiality}
P.~Doyle, J.~I. Lane, J.~J. Theeuwes, and L.~V. Zayatz, ``Confidentiality,
  disclosure, and data acces: theory and practical applications for statistical
  agencies,'' 2001.

\bibitem{yu2008privacy}
S.~Yu, G.~Fung, R.~Rosales, S.~Krishnan, R.~B. Rao, C.~Dehing-Oberije, and
  P.~Lambin, ``Privacy-preserving cox regression for survival analysis,'' in
  \emph{Proceedings of the 14th ACM SIGKDD international conference on
  Knowledge discovery and data mining}.\hskip 1em plus 0.5em minus 0.4em\relax
  ACM, 2008, pp. 1034--1042.

\bibitem{fung2008privacy}
G.~Fung, S.~Yu, C.~Dehing-Oberije, D.~De~Ruysscher, P.~Lambin, S.~Krishnan, and
  R.~R. Bharat, ``Privacy-preserving predictive models for lung cancer survival
  analysis,'' \emph{Practical Privacy-Preserving Data Mining}, p.~40, 2008.

\bibitem{chen2011privacy}
T.~Chen and S.~Zhong, ``Privacy-preserving models for comparing survival curves
  using the logrank test,'' \emph{Computer methods and programs in
  biomedicine}, vol. 104, no.~2, pp. 249--253, 2011.

\bibitem{kleinbaum1996survival}
D.~G. Kleinbaum and M.~Klein, \emph{Survival analysis}.\hskip 1em plus 0.5em
  minus 0.4em\relax Springer, 1996.

\bibitem{kleinbaum2012parametric}
------, ``Parametric survival models,'' in \emph{Survival analysis}.\hskip 1em
  plus 0.5em minus 0.4em\relax Springer, 2012, pp. 289--361.

\bibitem{ypma1995historical}
T.~J. Ypma, ``Historical development of the newton-raphson method,'' \emph{SIAM
  review}, vol.~37, no.~4, pp. 531--551, 1995.

\bibitem{dwork2013algorithmic}
C.~Dwork and A.~Roth, ``The algorithmic foundations of differential privacy,''
  \emph{Theoretical Computer Science}, vol.~9, no. 3-4, pp. 211--407, 2013.

\bibitem{kyle2006prevalence}
R.~A. Kyle, T.~M. Therneau, S.~V. Rajkumar, D.~R. Larson, M.~F. Plevak, J.~R.
  Offord, A.~Dispenzieri, J.~A. Katzmann, and L.~J. Melton~III, ``Prevalence of
  monoclonal gammopathy of undetermined significance,'' \emph{New England
  Journal of Medicine}, vol. 354, no.~13, pp. 1362--1369, 2006.

\bibitem{dispenzieri2012use}
A.~Dispenzieri, J.~A. Katzmann, R.~A. Kyle, D.~R. Larson, T.~M. Therneau, C.~L.
  Colby, R.~J. Clark, G.~P. Mead, S.~Kumar, L.~J. Melton \emph{et~al.}, ``Use
  of nonclonal serum immunoglobulin free light chains to predict overall
  survival in the general population,'' in \emph{Mayo Clinic Proceedings},
  vol.~87, no.~6.\hskip 1em plus 0.5em minus 0.4em\relax Elsevier, 2012, pp.
  517--523.

\bibitem{irgens2000medical}
L.~M. Irgens, ``The medical birth registry of norway. epidemiological research
  and surveillance throughout 30 years,'' \emph{Acta obstetricia et
  gynecologica Scandinavica}, vol.~79, no.~6, pp. 435--439, 2000.

\bibitem{wichert2008simple}
L.~Wichert and R.~A. Wilke, ``Simple non-parametric estimators for unemployment
  duration analysis,'' \emph{Journal of the Royal Statistical Society: Series C
  (Applied Statistics)}, vol.~57, no.~1, pp. 117--126, 2008.

\bibitem{smith2008efficient}
A.~Smith, ``Efficient, differentially private point estimators,'' \emph{arXiv
  preprint arXiv:0809.4794}, 2008.

\bibitem{dwork2010differential}
C.~Dwork and A.~Smith, ``Differential privacy for statistics: What we know and
  what we want to learn,'' \emph{Journal of Privacy and Confidentiality},
  vol.~1, no.~2, p.~2, 2010.

\end{thebibliography}

\end{document}